\documentclass[twocolumn,preprintnumbers,amsmath,amssymb]{revtex4}

\usepackage{graphicx}  
\usepackage{epstopdf}
\begin{document}

\title{Resonant plasmonic terahertz detection in vertical  graphene-base  hot-electron transistors}

\author{V.~Ryzhii$^{1,2}$, T.~Otsuji$^{1}$,  M.~Ryzhii$^{3}$, V. Mitin$^4$,  and M. S. Shur$^5$
} 
\affiliation{$^1$Research Institute of Electrical Communication, Tohoku University, Sendai 980-8577, Japan\\
$^2$ Center for Photonics and Infrared Engineering,
Bauman Moscow State Technical University and Institute of Ultra High
Frequency Semiconductor Electronics of RAS, Moscow 111005, Russia\\
$^3$Department of Computer Science and Engineering, University of Aizu, Aizu-Wakamatsu 965-8580, Japan\\
$^4$ Department of Electrical Engineering, University at Buffalo, SUNY, Buffalo, New York 1460-1920, USA\\
$^5$ Department of Electrical, Computer, and System Engineering and Physics, Applied Physics, and Astronomy,
Rensselaer Polytechnic Institute, Troy, New York 12180, USA}

\begin{abstract}
We analyze dynamic properties of vertical  graphene-base  hot-electron
transistors (GB-HETs) and consider their operation as detectors of terahertz (THz) radiation
using the developed device model.
The GB-HET  model accounts for the tunneling electron injection  from the emitter, electron propagation 
across the barrier layers with  the partial capture into the
GB,  and the self-consistent oscillations of the electric potential and the hole density 
in the GB (plasma oscillations), as well as the quantum capacitance and the electron transit-time  effects.
Using the proposed device model, we calculate the responsivity of GB-HETs operating as THz detectors
as a function of the signal frequency, applied bias voltages,  and the structural parameters.
The inclusion of the plasmonic  effect leads to the possibility of the HET-GBT operation at the frequencies
significantly exceeding those limited by the characteristic RC-time.   It is  found that the responsivity 
of GB-HETs with a sufficiently perfect GB exhibits sharp resonant maxima in the THz range of frequencies 
associated with the excitation of plasma oscillations. The positions of these maxima are controlled 
by the applied bias voltages. 
The GB-HETs can compete with and even surpass other plasmonic THz detectors.
\end{abstract}

\maketitle

\section{Introduction}

Recently,  vertical hot-electron    transistors (HETs) 
 with the graphene  base (GB) 
  and the bulk emitter and collector 
separated from the  base by the barrier layers - the hot-electron graphene-base transistors (GB-HETs) -
 made of  SiO$_2$ and
Al$_2$O$_3$ were fabricated and studied~\cite{1,2,3,4}. 
These HETs are fairly promising devices despite their modest characteristics at the present. 
Similar devices can be based of GL heterostructures with  
the hBN, WS$_2$, and other barrier layers~\cite{5,6,7,8}. 
 The history   of different versions of
HETs, including those  with the  thin metal base and the quantum-well (QW) base, in which the carriers 
are generated from impurities or induced by the applied voltages, 
as well as HETs with resonant-tunneling emitter, is rather long 
(see, for example, 
Refs.~\cite{9,10,11,12,13,14,15,16}). 
Figure~1 shows schematically the GB-HET structure and its band diagrams.
Depending on the GB doping, the base-emitter and collector-base voltages, $V_B$ and $V_C$,
and the thicknesses of barrier layers separating the base from the emitter and collector, $W_E$
and $W_C$, respectively,
the GB can be filled either with electrons or holes [compare Figs.~1(b) and 1(C)].

In GB-HETs (as in HETs) , a significant fraction of the electrons injected from the top n-type emitter 
contact due to tunneling through the barrier top
 crosses the GB and reaches the collector, while the fraction of the electrons captured into the
 GB can be rather small.
The tunneling  electrons  create the emitter-collector current $J_{C}$. The variations of  
the base-emitter voltage $V_B$ result in
the variation~of~$J_{C}$. 
\begin{figure}
\begin{center}
\includegraphics[width=7.9cm]{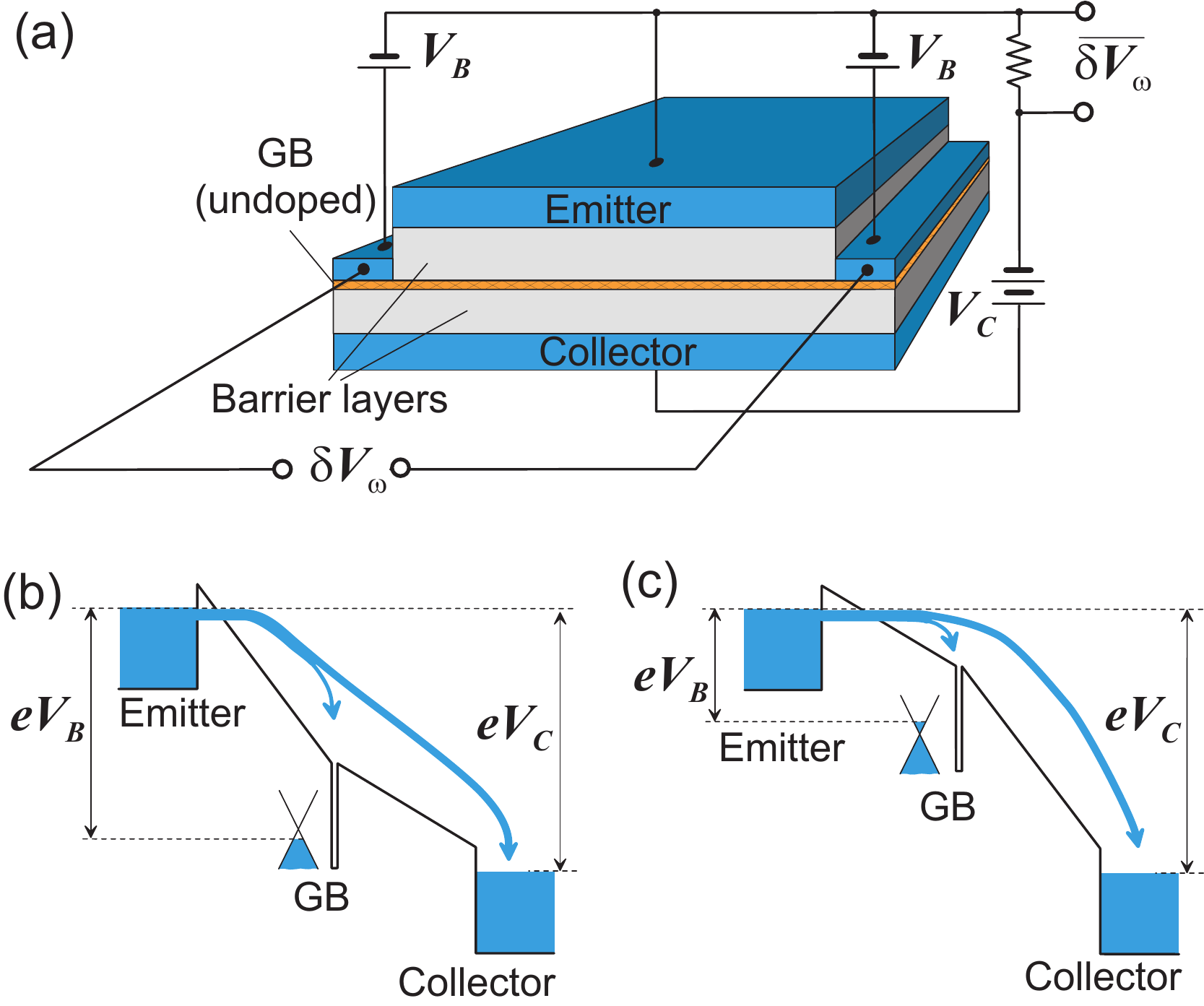}
\end{center}
\caption{
(a) Structure of a GB-HET with undoped GB   and GBT -HET band diagrams (potential profile in the direction 
perpendicular to the GB plane) at different relations between
$V_B$ and $V_C$: (b) with GB  filled with holes and (c) with GB filled with electrons. Arrows show  
propagation of electrons across the barrier layers and capture
of some portion of electrons into GB. }
\label{f1}
\end{figure}

In the case of the GB with a two-dimensional electron gas (2DEG), the GB-HET transistors can also be referred to as the N-n-N heterostructure HETs with the GB of n-type. In another case, when the GB comprises a two-dimensional hole (2DHG), the GB-HET transistors can also be called the N-p-N heterostructure bipolar transistors (HBTs) with the GB of p-type.

The electrons captured into the GB support the emitter-base current $J_{B}$.  The transistor gain 
in the common-emitter configuration is equal to $g_0 = J_{C}/J_{B} = (1 - p)/p$, where
$p$ is   the average capture probability of the capture of electrons into the GB during their
transit.
 One of the main potential advantages of GB-HETs is the high-speed operation associated 
with the  combination of a short transit time (due to the vertical structure), a high gain $g_0$ 
(because of a low probability, $p$, of the hot electron capture into the GB), and a low GB resistance  
(owing to a high mobility of the carriers in graphene). As shown recently~\cite{17},
$p$ in the graphene heterostructures with proper inter-graphene  barrier layers can be very small.
Very small values of $p$ in GB-HETs
can lead to a fairly high  gain. The use of the undoped base with the induced carriers, leads to 
the exclusion  of the scattering of the hot electrons crossing the GB with the impurities
and to an increase of the mobility of the carriers localized in the GB.

By analogy with HETs with the QW-base made of the standard materials~\cite{18,19}, 
GB-HETs can exhibit resonant response
to the incoming signals associated with the excitation of the plasma oscillations. 
The resonant plasma frequencies are determined by the characteristic plasma wave velocity $s$
(which increases with  the carrier density $\Sigma_0$) and the lateral sizes of the
GB $2L$, while the quality factor of the plasma resonances is mainly limited by the carrier momentum 
relaxation time $\tau$ associated with the scattering on impurities, various imperfections, and phonons.  
Due to the specific features of the carrier statistics and dynamics in the graphene layers~\cite{20},
the  plasma velocity $s > v_W$, where $v_W \simeq 10^8$~cm/s is the characteristic velocity of 
the Dirac energy spectrum, and $s$ can markedly exceed that in the QW heterostructures.
This promotes the realization of  with the plasmonic resonances in the terahertz (THz) range even in the GB-HETs
with fairly large lateral sizes. The possibility of achieving the elevated carrier mobilities in 
the GB
 can enable sharp  plasmonic resonances in GB-HETs at room temperatures or above.

The application of the  ac voltages  $\pm \delta V_{\omega}/2$ at the signal frequency $\omega$, associated with
the incoming radiation (received by an antenna)
 to the side   contacts to the GB, results in the variation 
of  the local ac potential difference $\delta \varphi_{\omega} = \delta \varphi_{\omega}(x) $ (the axis $x$ 
is directed in the GB plane) between the GB and the emitter contact. 
The emitter, base,  and collector  current densities $j_E$, $j_B$, and $j_C$  include  
 the dc components $j_{E,0}$, $j_{B,0}$, and   $j_{C,0}$ (determined by the applied bias voltages $V_{B}$
 and $V_{C}$), the ac components $\delta j_{E,\omega}$,  $\delta j_{B,\omega}$, and $\delta j_{C,\omega}$ 
 (proportional to $\delta V_{\omega}$, and (due to 
the  nonlinear dependence of the tunneling injection current  on the local potential difference between 
the GB and the emitter)
 the rectified dc components  
 $\overline {\delta j_{E,\omega}}$, $\overline {\delta j_{B,\omega}}$, and $\overline {\delta j_{C,\
 }}$ (proportional to $|\delta V_{\omega}|^2$, i.e., to the intensity of the incoming radiation received 
 by an antenna. The net rectified current $\overline {\delta J_{C,\omega}}$
 can serve as  the output current
in the  GB-HETs radiation  detectors.
The detector responsivity $R_{\omega} \propto \overline {\delta J_{C,\omega}}/|\delta V_{\omega}|^2$. 
 Due to the possibility of the plasmonic resonances, the rectified component (the detector output signal) 
 can be resonantly large, similar to that in the HET detector~\cite{21}
 and other plasmonic THz detectors using  different transistor structures, including those incorporating graphene~\cite{22,23,24,25,26,27,28,29,30,31,32,33,34}.

In this paper, develop the GB-HET device model  and evaluate the GB-HET characteristics as  a radiation detector of  
radiation, in particular, in the THz range of signal frequencies.

 \section{Device model and related equations}
 
We  consider  a GB-HET
with  the  highly conducting tunneling emitter and collector of the n-type  and the undoped
 GB (as shown in Fig.~1) with a 2DHG induced by the applied bias voltages
 creating the N-p-N structure with  the band diagram shown in Fig.~1(b). 
 The GB-HET device model accounts for  the tunneling injection of  hot-electrons from the emitter
 to the barrier layer(above the barrier top), their propagation across the barrier layers,
 partial capture of hot-electrons into the
GB,  and the excitation of the self-consistent oscillations of the electric potential and the hole density 
in the GB (plasma oscillations).
The quantum capacitance and the electron transit-time  effects are also taken into account.

We assume that 
 the bias voltages
$V_{B}$ and $V_{C}$
are applied between the GL-base and the emitter and between the collector and the emitter 
 contacts,
 respectively. 
In the framework of the gradual channel approximation~\cite{35}, which is valid if $W_E, W_C \ll 2L$, 
the density of the two-dimensional hole gas (2DHG)$\Sigma = \Sigma(x,t)$  in the  GL-base and its local 
potential $\varphi = \varphi(x,t)$ (counted from the potential of the emitter) are related to each other as

 \begin{equation}\label{eq1}
\Sigma = \frac{\kappa}{4\pi\,e}\biggl(\frac{\varphi - \mu/e - V_{bi}}{W_E} +
\frac{ \varphi - \mu/e -  V_{bi} - V_C}{W_C}\biggr),
\end{equation} 
 where  $\kappa$ is the dielectric constant of these layers, $e$ is the hole charge, $V_{bi}$ is 
 the built-in voltage between the contact material and an undoped GL,
 and $\mu$ is the 2DHG  Fermi energy in the GB.  
 In a degenerate 2DHG, $\mu  = \hbar\,v_W\sqrt{\pi\Sigma}$, where    $\hbar$ is the Planck constant.
The dependence of the right side of Eq.~(1) on the hole Fermi energy is interpreted as the effect of quantum 
capacitance~\cite{36,37}. 

The ac voltages 
 $\pm\delta V_{\omega}/2$ and $\omega$ are 
 applied between the side GB contacts connected with an antenna, so that
 the ac potential of the GB $\delta \varphi_{\omega} = \delta \varphi_{\omega}(x)$ obeys the following 
 (asymmetric) conditions: 
   
 \begin{equation}\label{eq2}
\delta \varphi_{\omega}|_{x = \pm L} = \pm \delta V_{\omega}/2.
\end{equation} 
 The side contacts can serve
 as the slot wave guide transforming the incoming THz radiation signals being received by an antenna 
 into the ac voltage [see Fig.1(a)].
 The ac component of the collector-emitter voltage $\delta V_{C,\omega}$ can also arise due to
 the ac potential drop across the load resistance. 
 However, in the GB-HET with the wiring  under 
 consideration, $\delta V_{C,\omega}$ can be disregarded providing that the GB-HET structure is
 symmetrical (see below).
 For GB-HETs with the degenerate 2DHG, the relation between the variations of the hole density and 
 the potential, $\delta \Sigma_{\omega}$ and $\delta \varphi_{\omega}$ (the ac components)
are, as follows from Eq.~(1),   can be expressed via the net  capacitance per unit area
$C = C_gC_{quant}/(C_{quant} + C_g)$, which accounts for the geometrical capacitance 
$C_g = (C_E + C_C) = (\kappa/4\pi)(W_E^{-1} +W_C^{-1})$ (with $C_E \propto W_E^{-1}$ and   $C_C \propto W_C^{-1}$ being the geometrical emitter and collector capacitances,
respectively) and the quantum capacitance~\cite{36,37} 
$C_{quant} = (2e^2\sqrt{\Sigma_0}/\sqrt{\pi}\hbar\,v_W) = (2e^2\mu_0/\pi\hbar^2v_W^2)$, where $\Sigma_0$
and $\mu_0$ are the pertinent dc values of the hole density and the Fermi energy.

\begin{equation}\label{eq3}
e\delta\Sigma_{\omega} = C\,\delta\varphi_{\omega}.
\end{equation}

Considering the electron tunneling from the emitter to the states above the top of the barrier
(through the triangular barrier), the emitter electron tunneling current density $j_E$ can be presented as:

 \begin{equation}\label{eq4} 
j_E = j_E^{t}
\exp\biggl(-\frac{F}{F_E}\biggr).
\end{equation}
Here 
$F =(a\sqrt{m}\Delta^{3/2}/e\hbar)$,
is  the characteristic tunneling field,  $\Delta$~is  the activation energy for electrons in the contact, $m$ is the effective electron mass in the barrier material, $a \sim 1$ is a numerical coefficient, 
$F_E = (\varphi  - V_{bi} - \mu/e)/W_E$ is the electric field in the emitter barrier, and 
$j_E^t$ is the maximum current density which can be provided the emitter contact, 
As follows from Eqs.~(2) and (3), the ac component and the rectified component
of the emitter tunneling current   $\delta\,j_{E,\omega}$  and $\overline {\delta\,j_{\omega}}$ 
(for $F \gg F_0$) are, respectively,  given by

 \begin{equation}\label{eq5} 
\delta j_{E,\omega} = j_{E,0} \frac{F}{F_{E,0}^2}\delta F_{E,\omega} = \sigma_E\delta F_{E,\omega}.
\end{equation}

 \begin{multline}\label{eq6} 
\overline {\delta j_{E,\omega}} \simeq 
\frac{j_{E,0}}{2}\biggl[\frac{1}{2}\biggl(\frac{F}{F_{E,0}}\biggr)^2 - \biggl(\frac{F}{F_{E,0}}\biggr)\biggr]\, \biggl|\frac{\delta F_{E,\omega}}{F_{E,0}}\biggr|^2\\
\simeq \frac{\sigma_EF}{4}\, \biggl|\frac{\delta F_{E,\omega}}{F_{E,0}}\biggr|^2.
\end{multline}
Here 
$j_{E,0} =  j_E^t
\exp(-F/F_{E,0}))$, $F_{E,0} = (V_{E} - V_{bi} - \mu_0/e)/W_E$, and 
$\sigma_E = j_0(F/F_{E,0}^2)$ 
are the emitter dc current density (in the absence of the ac signals), the dc electric field
in the emitter barrier, and the emitter differential conductance, respectively.
The dc hole Fermi energy in the GB $\mu_0$ obeys the following equation:
  
 \begin{equation}\label{eq7}
\mu_0^2 = \frac{\kappa\hbar^2v_W^2}{4e}\biggl[\biggl(V_{E} - V_{bi} - \frac{\mu_0}{e}\biggr)\biggl(\frac{1}{W_E} +  \frac{1}{W_C}\biggr)
- \frac{V_{C}}{W_C}\biggr].
\end{equation} 

The ac electric field components perpendicular to the GB plane  
in the emitter barrier, as well as in the collector barrier, are respectively given by

\begin{equation}\label{eq8} 
\delta F_{E,\omega} 
= \frac{C_{quant}}{(C_{quant} + C_g)}\frac{\delta \varphi_{\omega}}{W_E}, 
\end{equation}

\begin{equation}\label{eq9} 
\delta F_{C,\omega} 
=-\frac{C_{quant}}{(C_{quant} + C_g)}\frac{\delta \varphi_{\omega}}{W_C}.
\end{equation}

Using the Shockley-Ramo theorem, one can find the ac component the electron current densities
coming  to the base and the collector:

\begin{equation}\label{eq10} 
\delta j_{EB,\omega} = \sigma_{EB,\omega}\delta F_{E,\omega}, \qquad \delta j_{EC,\omega} = \sigma_{EC,\omega}\delta F_{E,\omega}.
\end{equation}

Here

\begin{multline}\label{eq11} 
\sigma_{EB,\omega} = 
\sigma_E\biggl[\frac{1}{W_E}\int_0^{W_E}dz e^{i\omega\,z/v}\\ - 
\frac{(1 - p)}{W_C}\int_{W_E}^{W_E + W_C}dz e^{i\omega\,z/v}\biggr]\\
= \frac{\sigma_E}{i\omega}\biggl[\frac{e^{i\omega\tau_E} - 1}{\tau_E}
- \frac{(1 - p)e^{i\omega\tau_E}(e^{i\omega\tau_C} - 1)}{\tau_C}\biggr],
\end{multline}

\begin{multline}\label{eq12} 
\sigma_{EC,\omega} = 
\sigma_E\frac{(1 - p)}{W_C}\int_{W_E}^{W_E + W_C}dz e^{i\omega\,z/v}\\
=  \frac{\sigma_E}{i\omega}\frac{(1 - p)e^{i\omega\tau_E}(e^{i\omega\tau_C} - 1)}{\tau_C}.
\end{multline}
where $v$ is the drift velocity of the hot electrons crossing the barriers above their tops
(which is assumed to be constant)
and $\tau_E = W_E/v$ and $\tau_C = W_C/v$ are the electron transit time across the emitter and collector barrier layers.
The axis $z$ is directed perpendicular to the GB plane. 
Since under the boundary conditions~(2), $\delta\varphi_{\omega}(x) = - \delta\varphi_{\omega}(-x)$
(see below) and, hence, $ \delta F_{E,\omega}(x) = -\delta F_{E,\omega}(x)$,
the net ac currents 

\begin{equation}\label{eq13} 
\delta J_{EB,\omega} = \int_{-L}^{L}dx\delta j_{EB,\omega} = 
 \sigma_{EB,\omega}\, \int_{-L}^{L}dx\delta F_{E,\omega} = 0,
\end{equation}

\begin{equation}\label{eq14} 
\delta J_{EC,\omega} = \int_{-L}^{L}dx\delta j_{EC,\omega}
= 
 \sigma_{EC,\omega}\, \int_{-L}^{L}dx\delta F_{E,\omega} = 0.
\end{equation}

Simultaneously for the  rectified components of the dc current densities from the emitter to the GB 
and from the emitter to the collector  one obtains

\begin{equation}\label{eq15} 
\overline{\delta j_{EB,\omega}} 
\simeq \frac{p\sigma_EF}{4} \biggl|\frac{\delta F_{E,\omega}}{F_{E,0}}\biggr|^2
= p\Gamma\, |\delta \varphi_{\omega}|^2,
\end{equation}

\begin{equation}\label{eq16} 
\overline{\delta j_{EC,\omega}}
\simeq \frac{(1 -p)\sigma_EF}{4} \biggl|\frac{\delta F_{E,\omega}}{F_{E,0}}\biggr|^2
= (1 - p)\Gamma\, |\delta \varphi_{\omega}|^2,
\end{equation}
where 
\begin{equation}\label{eq17} 
\Gamma = \frac{\sigma_EF}{4F_{E,0}^2W_E^2}\,\frac{C_{quant}^2}{(C_{quant} + C_g)^2}. 
\end{equation}
Consequently, the net rectified  components of the dc emittter-base and emitter-collector currents
are given by

\begin{multline}\label{eq18} 
\overline{\delta J_{EB,\omega}} = \int_{-L}^{L}dx\overline{\delta j_{EB,\omega}}
\simeq 
p\Gamma \int_{-L}^{L}dx|\delta \varphi_{\omega}|^2,
\end{multline}

\begin{multline}\label{eq19} 
\overline{\delta J_{EC,\omega}} =  \int_{-L}^{L}dx\overline{\delta j_{EC,\omega}}
\simeq 
(1 - p) \Gamma \int_{-L}^{L}dx|\delta \varphi_{\omega}|^2,
\end{multline}
respectively.

The ac hole current along the  GB is given by

\begin{equation}\label{eq20} 
\delta J_{BB,\omega} = - \sigma_{BB,\omega}\frac{d\delta\varphi_{\omega}}{d x}\biggl|_{x = L},
\end{equation}
where the lateral ac conductivity of the GB $\sigma_{BB,\omega}$ is given by (see, for example, Refs.~\cite{38,39,40,41,42}

\begin{equation}\label{eq21} 
\sigma_{BB,\omega} = \frac{ie^2\mu_0}{\pi\hbar(\omega + i/\tau)}.
\end{equation}

\section{Plasma oscillations and rectified current}

To calculate  the rectified current components  using Eqs.~(15) and  (16), one needs to find the spatial distributions of the ac potentials in the GB $\delta\varphi_{\omega} = \delta\varphi_{\omega}(x)$.

For this purpose we use the hydrodynamic equations for the hole transport along the GB~\cite{43,44}
(see also Refs.~\cite{21,22}) coupled with the Poisson equation solved using  the gradual channel approximation [i.e.,
using  Eq.~(1)]. 
Linearizing the hydrodynamic equations and Eq.~(1) and  taking into account that the ac component of the hole Fermi energy is expressed via the variation of their density and the potential, 
  we arrive at the following equation for the ac component of the GB potential $\delta \varphi_{\omega}$ (compare with Refs.~\cite{16,21,22} which should be solved with boundary conditions given by Eq.~(2):

\begin{equation}\label{eq22}
\frac{d^2\delta \varphi_{\omega}}{d\,x^2} +
\frac{(\omega + i\nu)(\omega + i\overline{\nu})}{s^2}
\delta \varphi_{\omega}
 = 0.
\end{equation}
Here   $s$ is
 the characteristic velocity of the plasma waves
in the gated graphene layers, which is  given by $s = \sqrt{e^2\Sigma_0/mC} \propto \Sigma_0^{1/4}$~\cite{18},  $m = \mu_0/v_W^2 \propto \sqrt{\Sigma_0}$ 
being the so-called "fictituous" 
effective hole (electron) mass in graphene layers~\cite{27},  $\overline{\nu} = (\sigma_{B,\omega}/W_EC_g)$ and $\nu = 1/\tau + \tilde{\nu}$, 
where $\tau$ is the hole momentum relaxation time in the 2DHG and
$\tilde{\nu} = \tilde{\nu}_{visc}  + \tilde{\nu}_{rad} $ is associated with  the contribution of the 2DHG 
viscosity to the damping (see for example, Ref.~\cite{21}) and 
with the radiation damping of the plasma
oscillations. 
The latter mechanism is associated with  the 
recoil that the holes in the GB   feel emitting radiation (the pertinent term  
in the force acting on  the holes is referred to as the Abraham-Lorentz
force or the radiation reaction force~\cite{46,47}).
Taking into account that the viscosity damping rate is proportional to the second spatial derivative, for  
the gated plasmons with the acoustic-like spectrum $\tilde{\nu}_{visc} = \zeta \omega^2/s^2$
($\zeta$ is the 2DHG viscosity),and  considering that $\tilde{\nu}_{rad} \propto (2e^2/3mc^3)\omega^2$~\cite{45,46,47}),
we put $\nu = 1/\tau  + \eta\omega^2)$, where $\eta$ 
is the pertinent damping parameter, and $c$ is the speed of light. If  the viscosity damping surpasses 
the radiation damping, one can set  
$\eta \simeq \zeta/s^2$.

 The characteristic plasma-wave velocity $s$  is determined by $\Sigma_0$ (as well as the thicknesses
 of the emitter and barrier layers $W_E$ and $W_C$)~\cite{20} and, hence, can be changed by the variations of the bias voltages $V_E$ and $V_B$. Due to this the characteristic plasmonic frequency
$\Omega =\pi\,s/L$ can be effectively controlled by these voltages.


Equation ~(22) with Eq.~(2) yield

\begin{equation}\label{eq23} 
\delta\varphi_{\omega}   =
\frac{\sin(\sqrt{(\omega + i\nu)(\omega + i\overline{\nu})})x/s] 
}
{\sin [\sqrt{(\omega + i\nu)(\omega + i\overline{\nu})}L/s] 
}\frac{\delta V_{\omega}}{2} .
\end{equation}
 One can see from Eq.~(23) that the spatial dependence of $\delta\varphi_{\omega}$
is rather complex. In particular, when $\omega$ approaches to
 $n\Omega$, where
$n = 1,2,3,...$, this distribution can be 
oscillatory with fairly high amplitude of the spatial oscillations when $\Omega \gg \nu$.

Substituting $\delta\varphi_{\omega}$ given by Eq.~(23) to Eqs.~(15) and (16), we arrive at
the following:

\begin{multline}\label{eq24} 
\overline{\delta j_{EB,\omega}}   \simeq p\Gamma\\
\times\biggl|\frac{\sin[\sqrt{(\omega + i\nu)(\omega + i\overline{\nu})}x/s]}{\sin [\sqrt{(\omega + i\nu)(\omega + i\overline{\nu})}L/s]}\biggr|^2\frac{|\delta V_{\omega}|^2}{4},
\end{multline}

\begin{multline}\label{eq25} 
\overline{\delta j_{EC,\omega}}   \simeq (1 -p)\Gamma\\
\times\biggl|\frac{\sin[\sqrt{(\omega + i\nu)(\omega + i\overline{\nu})}x/s]}{\sin [\sqrt{(\omega + i\nu)(\omega + i\overline{\nu})}L/s]}\biggr|^2\frac{|\delta V_{\omega}|^2}{4},
\end{multline}

\begin{multline}\label{eq26} 
\overline{\delta J_{EB,\omega}}   \simeq p\Gamma\\
\times\int_{-L}^Ldx\biggl|\frac{\sin[\sqrt{(\omega + i\nu)(\omega + i\overline{\nu})}x/s]}{\sin [\sqrt{(\omega + i\nu)(\omega + i\overline{\nu})}L/s]}\biggr|^2\frac{|\delta V_{\omega}|^2}{4},
\end{multline}

\begin{multline}\label{eq27} 
\overline{\delta J_{EC,\omega}}   \simeq (1 -p)\Gamma\\
\times\int_{-L}^Ldx\biggl|\frac{\sin[\sqrt{(\omega + i\nu)(\omega + i\overline{\nu})}x/s]}{\sin [\sqrt{(\omega + i\nu)(\omega + i\overline{\nu})}L/s]}\biggr|^2\frac{|\delta V_{\omega}|^2}{4}.
\end{multline}

As follows from Eqs.~(24) - (27), $\overline{\delta j_{EC,\omega}} =
[(1 - p)/p]\overline{\delta j_{EB,\omega}}$
and
$\overline{\delta J_{EC,\omega}} =
[(1 - p)/p]\overline{\delta J_{EB,\omega}}$ (as for the pertinent dc current in the absence of the THz signals), so that  $\overline{\delta J_{EC,\omega}} \gg
\overline{\delta J_{EB,\omega}}$.

Using Eqs.~(20) and (24), for the ac current between the based contacts we obtain

\begin{multline}\label{eq28}
\delta J_{BB,\omega} = - \frac{ie^2\mu_0\cot [\sqrt{(\omega + i\nu)(\omega + i\overline{\nu})}L/s]}{\pi\hbar\,s}\\
\times\sqrt{\frac{\omega + i\overline{\nu}}{\omega + i\nu}} \frac{\delta V_{\omega}}{2}
\end{multline}
Due to the symmetry of the GB-HET structure and the asymmetric spatial distribution
of $\delta\varphi_{\omega}$, there is no rectified component of the lateral current in the GB-base, 
i.e., $\overline {\delta J_{BB,\omega}} = 0$.

 At very low signal frequencies $\omega \ll \nu, |\overline{\nu}| \simeq 4\pi\,p\sigma_E/\kappa$, from Eqs.~(23) and (22) we obtain 

\begin{equation}\label{eq29} 
\delta\varphi_{\omega}   \simeq 
\frac{\sinh(\sqrt{\nu\overline{\nu}}x/s) 
}
{\sinh (\sqrt{\nu\overline{\nu}}L/s) 
}\frac{\delta V_{\omega}}{2},
\end{equation}

\begin{multline}\label{eq30} 
\overline{\delta J_{EC,\omega}}   \simeq (1 -p)\Gamma\,L
\biggl[\frac{\displaystyle\frac{\sinh(2\sqrt{\nu\overline{\nu}}L/s)}{(2\sqrt{\nu\overline{\nu}}L/s)} -\displaystyle 1}{\sinh^2(\sqrt{\nu\overline{\nu}}L/s)}\biggr]\frac{|\delta V_{\omega}|^2}{4}\\
\simeq (1 -p)\Gamma\,L\frac{|\delta V_{\omega}|^2}{6}.
\end{multline}

\begin{figure}
\begin{center}
\includegraphics[width=7.9cm]{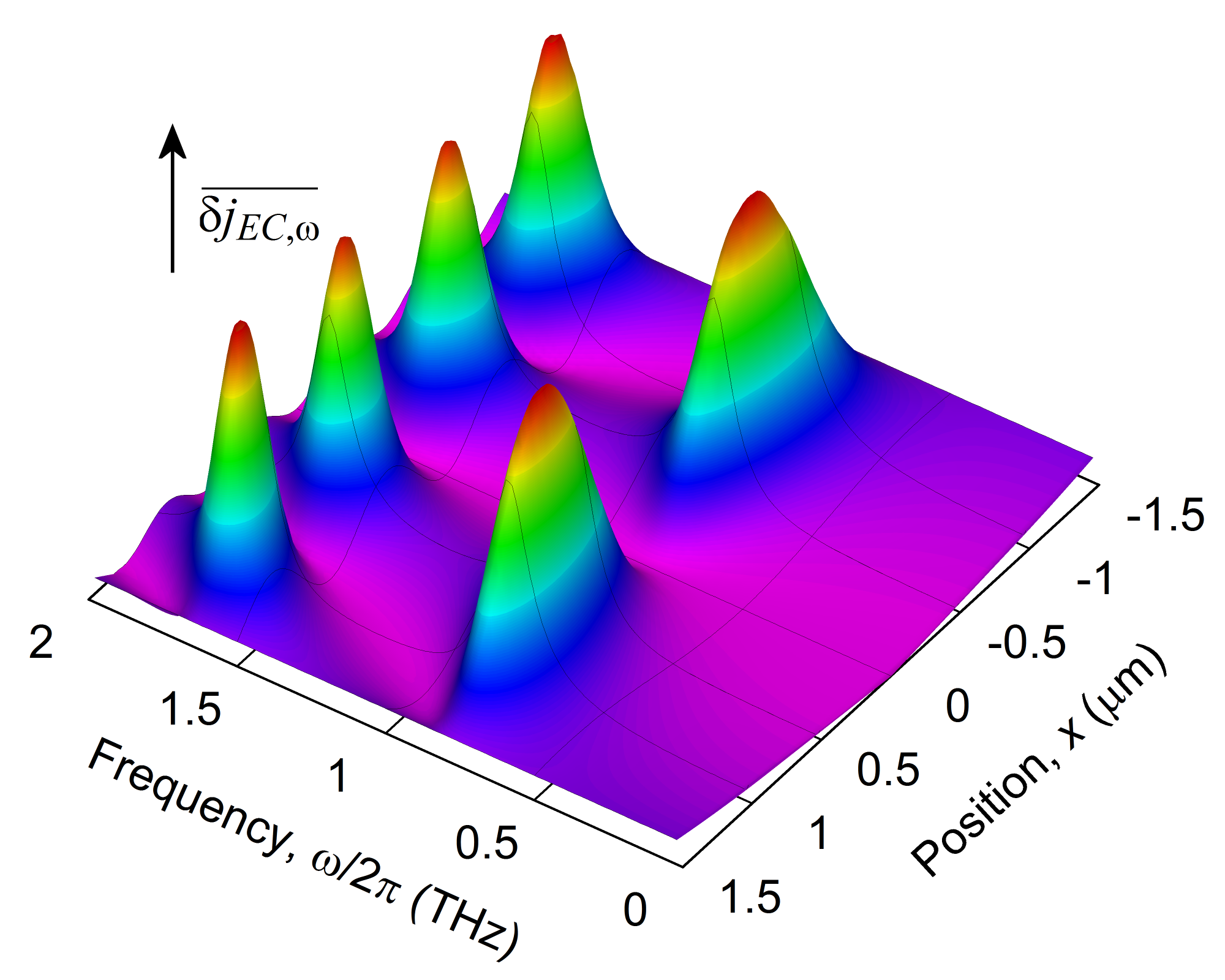}
\end{center}
\caption{
Spatial distributions of rectified component of 
 the emitter-collector current density
$\overline{\delta j_{EC,\omega}}$ at different frequencies $\omega$ ($\tau = 1$~ps, $s = 2.5\times 10^8$~cm/s, and $L = 1.5~\mu$m).}
\label{f2}
\end{figure}

Since in reality $|\overline{\nu}| \ll \nu$, 
there is an intermediate range of frequencies 
 $|\overline{\nu}| \ll \omega <\nu$. Assuming that $|\overline{\nu}| \ll \nu, \Omega^2/\omega < \nu$,
  from Eq.~(27) we arrive at

\begin{multline}\label{eq31} 
\overline{\delta J_{EC,\omega}}   \simeq (1 -p)\Gamma\,L
\biggl(\frac{s}{L}\sqrt{\frac{2}{\nu\omega}}\biggr)\frac{|\delta V_{\omega}|^2}{4}\\
\propto (1 - p)\sqrt{\frac{\Omega^2}{\nu\omega}}|\delta V_{\omega}|^2.
\end{multline}

If the characteristic frequency of the plasma oscillations $\Omega = \pi\,s/L \gg \nu$,
the ac GB potential amplitude  $|\delta \varphi_{\omega}|$  can markedly exceed 
the amplitude of the input ac signal $\delta V_{\omega}$ when the frequency is close to one of the resonant plasma frequencies $n\Omega$,   where $n = 1,2,3,...$ is the plasma resonance index.
In this case, the 
rectified emitter-collector current is pronouncedly stratified, i.e., its density  ${\overline \delta j_{EC,\omega}}$ is a nearly periodic function of the coordinate $x$.

Figure~2 shows the spatial distributions of the 
rectified emitter-collector current density (corresponding to the current stratification) calculated for different frequencies
using Eq.~(25).
It is assumed that $\tau = 1$~ps, $s = 2.5\times 10^8$~cm/s, and $L = 1.5~\mu$m.
As seen in Fig.~2, the spatial stratification  of the current  is rather pronounced when
$\omega$ is close to the plasma resonant frequencies (the frequencies   $\Omega = 5/6 \simeq 0.83$~THz and $2\Omega/2\pi = 5/3 \simeq 1.66$~THz): the current exhibits two streams centered at $|x|/L \simeq 0.5$ when $\omega \simeq 0.83$~THz  and four streams  centered at $|x|/L \simeq 0.25$
when
$\omega \simeq 1.66$~THz. At the frequencies far from the resonances, the spatial
current distribution becomes weakly  nonuniform.

\section{GB-HET detector responsivity}

\subsection{Current responsivity}

The rectified current $\overline {\delta J_{EC,\omega}}$ can be considered as the output signal
used for the detection of electromagnetic radiation (in particular, THz radiation).
The current detector  responsivity using this output signal is defined as

\begin{equation}\label{eq32} 
R_{\omega}   = \frac{\overline {\delta J_{EC,\omega}}H}{SI_{\omega}},
\end{equation}
where $H$ is the GB-HET lateral size in the direction along the contacts to the GB, $I_{\omega}$
is intensity of the incident radiation, and  
$S$ is the antenna aperture. The latter is given by $S = \lambda_{\omega}^2G/4\pi$~\cite{50}, where $G$  is the antenna gain,  
$\lambda_{\omega} = 2\pi\,c/\omega$ is the radiation wavelength, and $c$ is the speed of light in vacuum.
Taking into account that $I_{\omega} = c{\cal E}^2_{\omega}/8\pi$, where ${\cal E}$ 
is the radiation electric field in vacuum, and estimating $\delta V_{\omega}$ as $\delta V_{\omega} = \lambda{\cal E}/\pi$,
one can arrive at

\begin{equation}\label{eq33} 
|\delta V_{\omega}|^2 = \frac{8\lambda^2_{\omega}I_{\omega}}{\pi\,c} = \frac{32\pi\,c}{\omega^2} I_{\omega}.
\end{equation}

Considering Eqs.~(27), (32), and (33), for the current responsivity (in A/W units) we find 

\begin{equation}\label{eq34} 
R_{\omega}   = \frac{R}{L}\int_{-L}^Ldx
\biggl|\frac{\sin[\sqrt{(\omega + i\nu)(\omega + i\overline{\nu})}x/s]}
{\sin [\sqrt{(\omega + i\nu)(\omega + i\overline{\nu})}L/s]}\biggr|^2,
\end{equation}
where

\begin{equation}\label{eq35} 
R  \simeq \frac{8(1-p)\Gamma\,LH}{cG} = \rho\frac{LH}{W_E^2}
\end{equation}
 \begin{equation}\label{eq36} 
\rho = \frac{8(1-p)j_0}{cG}\biggl(\frac{F}{F_{E,0}^2}\biggr)^2.
\end{equation}

As follows from Eq.~(34), at relatively long hole  momentum relaxation times $\tau$,
the GB-HET responsivity exhibits a series of very sharp and high peaks.
It is instructive that in the GB-HETs under consideration, 
the height of the resonant peaks does not decrease with an increasing peak index (as in some other devices using the
plasmonic resonances). Although 
some decrease in the peaks height with the increasing resonance index attributed to the effect of viscosity
and radiative damping takes place,  the pertinent  effect is relatively weak (see below).

In the limiting cases $\omega \ll \nu, |\overline{\nu}|$ and 
$|\overline{\nu}| \ll \nu, \Omega^2/\omega < \nu$, corresponding to Eqs.~(25) and (26),  one obtains

\begin{equation}\label{eq37} 
R_{\omega} = R_0  \simeq \frac{2}{3}R
\end{equation}
and 
\begin{equation}\label{eq38} 
R_{\omega}   \simeq R\biggl(\frac{s}{L}\sqrt{\frac{2}{\nu\omega}}\biggr)
=R\sqrt{\frac{2\Omega^2}{\pi^2\nu\omega}} < R,
\end{equation}
respectively. The quantities $R$ and $R_{\omega}$ depend on the geometrical and quantum capacitances and, therefore,
on the barrier layers thicknesses:

\begin{multline}\label{eq39}
R_{\omega} \propto R \propto  \frac{1}{W_E^2}\frac{C^2_{quant}}{(C_{quant} + C_g)^2}\\
 = \frac{1}{W_E^2}\biggl[1 + \frac{\kappa}{ 4\pi\, C_{quant}(W_E^{-1} + W_C^{-1})}\biggr]^{-2}.
\end{multline}

For the doping level of the emitter contact $N_D = 10^{18}$~cm$^{-3}$, assuming that the thermal electron velocity $v_T = 10^7$~cm/s, we obtain $j_E^t =1.6\times 10^6$~A/cm$^2$.
Setting $\Delta = 0.2$~eV and the effective mass in the barrier layer $m \simeq 2.5\times 10^{-28}$~g 
, we arrive at $F \sim 2\times 10^6$~V/cm. Setting also $F_{E,0} = 5\times 10^5$~V/cm
(to provide the hole density in the GB about of $\Sigma = 10^{12}$~cm$^{-2}$), we find $j_{E,0} \simeq 2.9\times 10^{4}$~A/cm$^2$ and 
$\sigma_E \simeq 0.23$~Ohm$^{-1}$cm$^{-1}$. 
At these parameters, one obtains also $C_{quant} \simeq  2.6\times 10^6$~cm$^{-1}$.
Using these data and setting in addition $p \ll 1 $, $W_E W_C/(W_E + W_C) = 5$~nm, $\kappa = 4$, and $G = 1.5$, from Eq.~(36) we obtain
$\rho \simeq 2\times 10^{-4}$~A/W. If $L = 1.5~\mu$m and $H = 10~\mu$m, Eqs.~(35) - (36)
yield $R \simeq 30$~A/W and $R_0 \simeq 20$~A/W. These parameters are also used in the estimates below.

In the case of high quality factor of the plasma resonances $\Omega/\nu$, 
the quantities $\overline{\delta J_{EB,\omega}}$ and  $\overline{\delta J_{EC,\omega}}$ 
and, hence, the responsivity as functions of the signal frequency $\omega$ described by Eqs.~(27) and (34) exhibit sharp peaks at $\omega \simeq
n\Omega$,  attributed to the resonant excitation of the plasma oscillations
(standing plasma waves). The peak width is primarily determined by the frequency $\nu$.
Indeed, for  $\Omega/\nu \gg 1$     Eq.~(34) yields 

\begin{equation}\label{eq40} 
{\rm max} R_{\omega} \simeq R_{\Omega} \simeq  \frac{3R_0 }{2}\biggl(\frac{2\Omega}{\pi\nu}\biggr)^2 >> R_0.
\end{equation}
Using the above estimate for $R_0$,  the peak values of the responsivity  at $\Omega = 5/3$~THz
and $\tau = 1$~ps is approximately equal to max$R_{\omega} \simeq R_{\Omega} \simeq 1.33\times 10^3$~A/W.

Figure ~3 shows the dependence of the normalized GB-HET current responsivity 
calculated using Eq.~(34)
 for several sets of parameters.
As seen, the positions of the responsivity peaks shift toward higher frequencies when the plasma
frequency increases, i.e., when $s$ increases and/or $L$ decreases because $\Omega \propto s/L$ (compare the curves "1"
and "2"). A shortening of the momentum relaxation time $\tau$ leads to a smearing of the peaks (compare the curve "3"
corresponding to $\tau = 1$~ps and the curve "4" corresponding to $\tau = 0.5$~ps).
Figure~4 shows the lowering and broadening of the resonance peaks of the GB-HET current responsivity
with the decreasing  momentum relaxation time $\tau$ described by Eq.~(34). As seen, at $\tau < 0.3 - 0.4$~ps, the responsivity peaks vanish, while at $\tau = 1$~ps they are fairly sharp and high..

\begin{figure}
\begin{center}
\includegraphics[width=7.0cm]{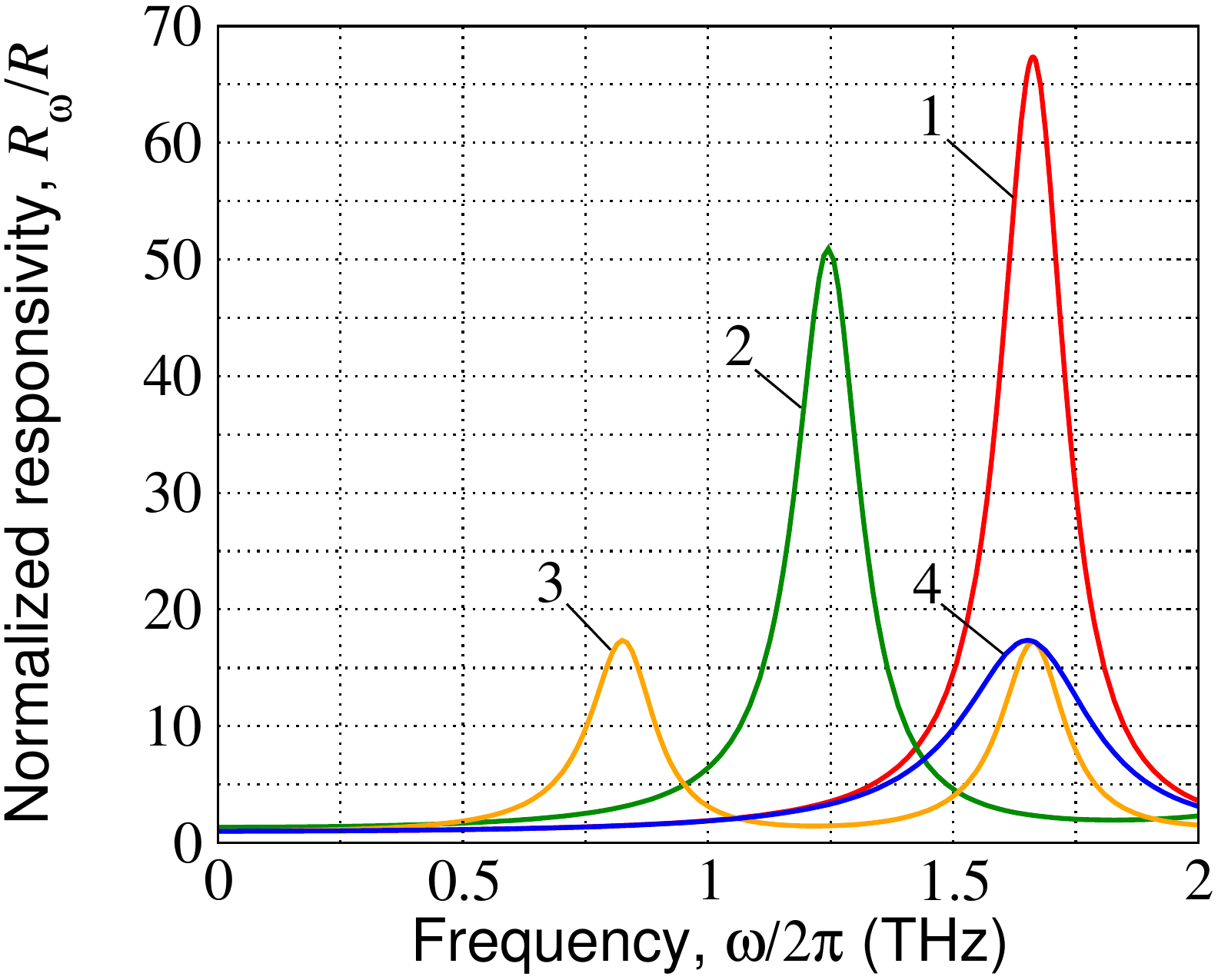}
\end{center}
\caption{
Normalized respoinsivity versus signal frequency for GB-HETs with different parameters:\\
1 - $\tau = 1$~ps, $s = 5\times 10^8$~cm/s, $L = 1.5~\mu$m;\\
2 - $\tau = 1$~ps, $s = 5\times 10^8$~cm/s, $L = 2.0~\mu$m;\\
3 - $\tau = 1$~ps, $s = 2.5\times 10^8$~cm/s, $L = 1.5~\mu$m;\\
4 - $\tau = 0.5$~ps, $s = 5\times 10^8$~cm/s, $L = 1.5~\mu$m.\\
These parameters correspond to $\Omega/2\pi =$ 5/3, 5/4, 5/6, and 5/3 THz, respectively.}
\label{f3}
\end{figure}

\begin{figure}
\begin{center}
\includegraphics[width=7.9cm]{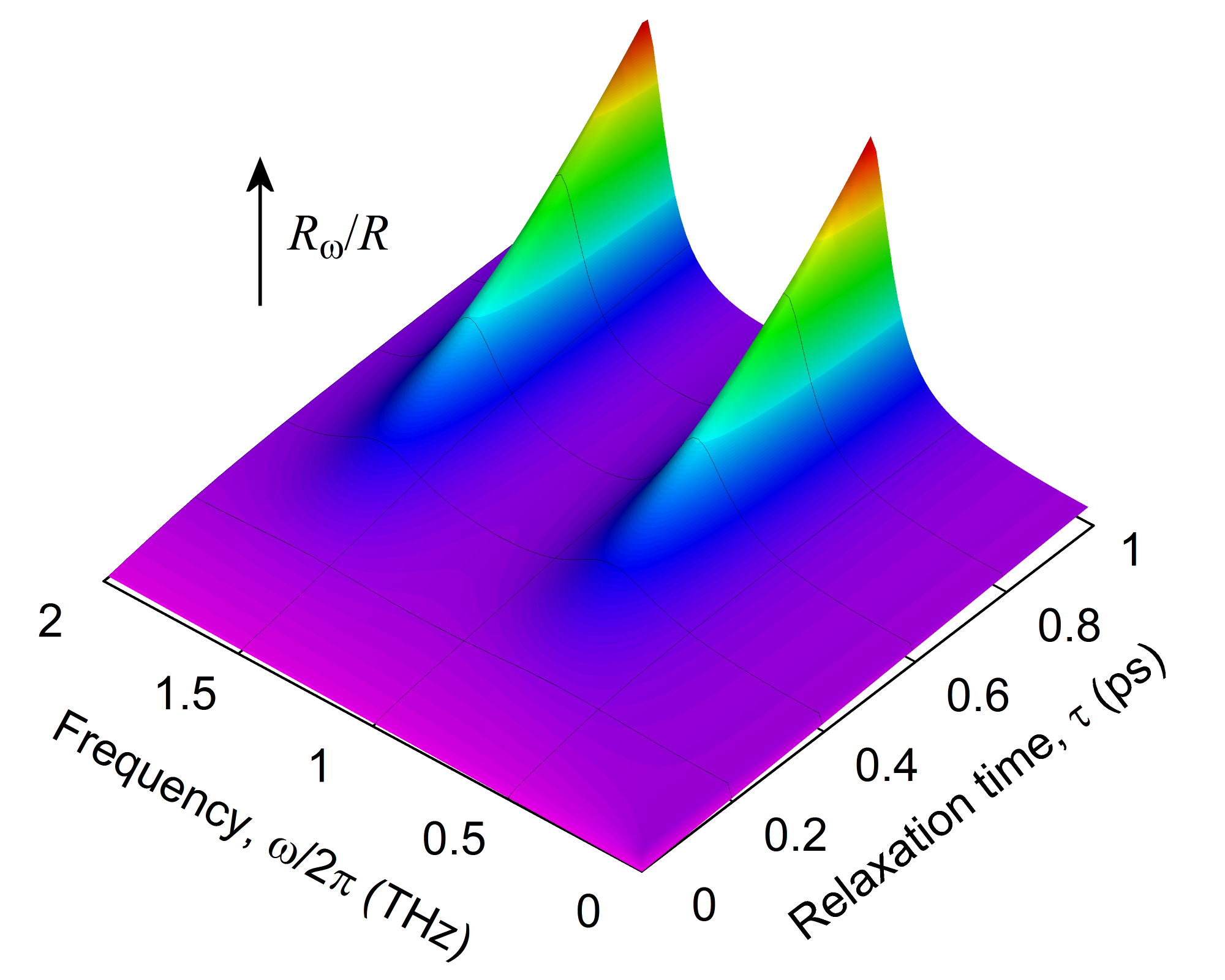}
\end{center}
\caption{
Normalized respoinsivity as a function of signal frequency and momentum relaxation time
($s = 2.5\times 10^8$~cm/s and $L = 1.5~\mu$m, i.e., $\Omega/2\pi = 5/6$~THz).}
\label{f3}
\end{figure} 
  \subsection{Voltage responsivity}

  
The variation of the dc current component  $\overline {\delta J_{EC,\omega}}$ cause by the ac signals results in a change
of the voltage drop, $\overline {\delta V_{\omega}} = - \delta V_{C,0}$, across the load resistor
in the collector circuit (see Fig.~1).
Considering that this leads to an extra variation of the dc emitter and collector currents
$\overline{\delta J_{E,0}} = [\sigma_E C_g/(C_{quant} + C_g)(W_E + W_C)]\delta V_{C,0}$ and 
 $\overline{\delta J_{EC,0}} = [(1 - p)\sigma_E C_g/(C_{quant} + C_g)]\delta V_{C,0}$  because of its dependence on the collector contact dc potential. The latter  dependence is essentially
 associated with the effect of quantum capacitance.
Taking this into account, for the voltage responsivity of the GB-HET under consideration
$R_{\omega}^{V} = \overline {\delta V_{\omega}}H/SI_{\omega} = \delta V_{C,0}H/SI_{\omega}$ we obtain

\begin{equation}\label{eq41} 
R_{\omega}^V   = \frac{R^V}{L} \int_{-L}^Ldx
\biggl|\frac{\sin[\sqrt{(\omega + i\nu)(\omega + i\overline{\nu})}x/s]}
{\sin [\sqrt{(\omega + i\nu)(\omega + i\overline{\nu})}L/s]}\biggr|^2.
\end{equation}
Here

\begin{equation}\label{eq42} 
R^V   = R \frac{r_C}{\biggl[1 + \displaystyle\frac{C_g}{(C_{quant} + C_g)}\frac{2LH}{(W_E + W_C)}r_C\sigma_E\biggr]},
\end{equation}
where $r_C$ is the load resistance.
It is instructive that due the the absence of the ac collector current (i.e., the ac current through
the load resistor), associated with the GB-HET structure symmetry and the asymmetry of the applied ac signal voltage and the ac potential spatial distribution along the GB) 
the RC-factor of the voltage responsivity is independent of the signal frequency.

For  $W_E = W_C = W$, Eqs.~(38) and (40) yield

\begin{equation}\label{eq43} 
R_{\omega}^V   = R_{\omega}\frac{r_C}{\biggl[1 + \displaystyle\frac{\sigma_Er_C(LH/W^2)W}{(1 + 2\pi\,C_{quant}W/\kappa )}\biggr]}
\end{equation}

At $r_C \ll (1 + 2\pi C_{quant}W/\kappa)(W/LH\sigma_E) = {\overline r_C}$, Eq.~(43) yields
the obvious formulas:

\begin{equation}\label{eq44} 
R_{0}^V = R_0 r_C
\end{equation}
at low frequencies, and

\begin{equation}\label{eq45} 
R_{\Omega}^V = \frac{3R_0}{2}\biggl(\frac{2\Omega}{\pi\nu}\biggr)^2 r_C
\end{equation}
at the plasma resonance $\omega = \Omega$.

At $r_C > (1 + 2\pi C_{quant}W/\kappa)(W/LH\sigma_E) = {\overline r_C}$, from Eq.~(43) we obtain, respectively,

\begin{equation}\label{eq46} 
 R_{\Omega}^V   \simeq R_0
\frac{(W^2/LH)}{\sigma_E\,W}\biggl(1 + \frac{2\pi\,C_{quant}W}{\kappa}\biggr).
\end{equation}
and

\begin{equation}\label{eq47} 
 R_{\Omega}^V   \simeq \frac{3R_0}{2}\biggl(\frac{2\Omega}{\pi\nu}\biggr)^2
\frac{(W^2/LH)}{\sigma_E\,W}\biggl(1 + \frac{2\pi\,C_{quant}W}{\kappa}\biggr).
\end{equation}
For 
$\tau = 1$~ps, $s = 5\times 10^8$cm/s, and $L = 1.5~\mu$m (as for the curve "1" in Fig.~3), so that $\Omega = 5/3$~THz, as well as $\sigma_E = 0.23$~A/V cm$, C_{quant} = 2.6\times 10^6$~cm$^{-1}$ (see the estimate in the previous subsection), and $\kappa = 4$ from Eq.(47) we obtain the following estimate:
$R_{\Omega}^V   \simeq 2\times 10^5 {\rm V/W}$.
For the above parameters one obtains  ${\overline r_C} \simeq 150$~Ohm.
Even at smaller $r_C$, the voltage responsivity can be fairly large. Setting $r_C = 5 - 10$~Ohm,
we obtain $R_0^V \simeq 100 - 200$~V/W   and $R_{\Omega}^V   \simeq (6.6 - 13.2)\times 10^3$~V/W,
respectively.


 Considering Eqs.~(42) and (39), we find the following dependences of the current and voltage responsivities $R_{\omega}$ and $R_{\omega}^V$ on the emitter and collector barriers thickness $W$ (at not too large $r_C$):

\begin{equation}\label{eq48}
R_{\omega}^V  = R_{\omega}r_C  \propto  
\frac{4\pi^2C_{quant}^2}{\kappa^2(1 + 2\pi\,C_{quant}W/\kappa)^2}.
\end{equation}
According to Eq.~(48), $R_{\omega}$ and $R_{\omega}^V$ markedly decrease in the range
$W > \kappa/2\pi C_{quant} \simeq 10$~nm.

\section{Discussion}
As follows from Eqs.~(34) and(40), an increase in $\nu$ with increasing frequency $\omega$ due to the 
reinforcement of the plasma oscillation damping associated with the viscosity and the radiative damping, might lead to the gradual
lowering of the resonant peaks with their index $n$. 
However, our estimates show that the contribution of
these two mechanisms to the net damping is small compared to the  damping associated with the hole momentum relaxation  (collisional damping).
Indeed, disregarding the radiative damping and assuming
the 2DHG viscosity  to be $\zeta = 10$~cm$^2$/s (i.e., smaller than in the standard 2DEG and 2DHG in the GaAs 
based heterostructures~\cite{21,49,50}) and $s = (2.5 - 5)\times 10^8$~cm/s, 
we obtain $\eta \simeq (4 - 16)\times 10^{-17}$~s. Hence, in the frequency range $\omega/2\pi \leq 2$~THz (as in Fig.~2), we find
${\tilde\nu}_{visc} = \eta\omega^2 \leq (6.3 - 25.3)\times 10^9$~s$^{-1}$ $ \ll 1/\tau$. Therefore, the heights of the responsivity peaks
in Fig.~2 in the curve "3" at $\omega \simeq 0.8$ and 1.6~THz are virtually equal. 
However, the peaks corresponding to higher resonances with the frequencies
in the range 5 - 10~THz can be markedly  lowered and smeared, because in this range
${\tilde\nu}_{visc}$ can become comparable with $1/\tau$. For example, for the same values of $\eta$, $\tau = 1$~ps,
and $\omega/2\pi$ = 5 - 10~THz, we obtain ${\tilde\nu}_{visc}\tau \simeq 0.16 - 0.64$.

 Equation~(42) describes the saturation of  the voltage responsivity $R_{\Omega}^V$ peak value with increasing load resistance $r_C$. This  is associated with the effect of the voltage drop across the load on the potential drop between the emitter and the base and, hence, the hole Fermi
 energy in the GB,  which determine the injection current.
Such an effect is due to the finite value of the GB quantum capacitance (see, Refs.~\cite{12,13}) - 
if $ C_{quant}$ tends to infinity, the emitter-base voltage becomes independent
of $r_C$, and the saturation of the $R_{\Omega}^V - r_C$ dependence vanishes.

The THz detectors using a similar operation principle and InP double heterojunction bipolar
transistors (DHBTs) were recently fabricated and studied experimentally~\cite{51,52,53}.
The estimated responsivities of the DHBTs in question 
for the  non-resonant detection regime are somewhat smaller but of the same order of magnitude than those given by Eq.~(43) and the pertinent estimates.
However, the experimental values of the responsivity are much smaller than the values predicted above for the resonant detection [see Eqs.~(44) and (46) and the estimates based on these equations].
Apart from the parasitic effects and the absence of any spatial coupling antenna, this can be attributed to the doping of the base in the InP-DHBTs, which  inevitably leads to relatively a
shorter hole momentum relaxation time $\tau$ compared to that in the GB (where  the 2DHG is induced by the applied voltages). 
Possibly, a higher probability  of the hot electron capture into the
InGaAs base (that had a relatively large thickness of 28~nm) in the DHBTs  in comparison with 
 the GB-HETs
can be an additional factor.

The GB-HET current and voltage responsivities are determined by several  characteristics:
the characteristics of the tunneling emitter, geometrical characteristics of the GB-HET structure,
materials of the emitter as well as the emitter and collector barrier layers, and applied bias voltages. The diversity of these factors enables the optimization of the GB-HETs operating as resonant plasmonic detectors, in particular, an increase in the responsivity in comparison with the values obtained in the above estimates. The resonant plasmonic THz detectors can based on not only the GB-HET
structure shown in Fig.~1(a) (with an extra antenna connected to the GB side contacts),
but also based on lateral  structures with the GB contacts forming a periodic array.

  The comparison of the GB-HETs~\cite{1,2,3,4} and  InP-DHBTs~\cite{51,52,53} with the GB-HETs under consideration highlights the following advantages of the latter: (i) a longer momentum relaxation time of holes $\tau$ in the GB;
(ii) a higher plasma-wave velocity $s$ that enables higher resonant plasma frequencies; (iii)
a smaller capture probability of hot electrons into the GB and, consequently, larger (or even much larger) fraction of the hot electrons reaching the collector; 
(iv) coupling  the incoming THz signal to
the GB resulting in the absence of the ac current in the emitter-collector circuit and prevanting
 the RC effects usually hindering the high-frequency operation.  

\section{Conclusions}

We developed an analytical model for vertical heterostructure HETs with the GB of the p-type
sandwiched between the wide-gap emitter and collector layers and the N-type contacts. Using this model, we described the GB-HET
 dynamic properties 
and studied the GB-HET  operation as  detectors of THz radiation. The main features of the GB-HETs are 
high hole mobility in the GB,  low probability the capture of the hot electrons injected from the emitter and crossing the GB, and the absence of the collector ac current.
These features enable pronounced voltage-controlled plasmonic response of the GB-HETs to the incoming THz radiation,
high hot-electron injection efficiency, and  the elimination of the RC-limitations  leading to elevated the GB-HET
current and voltage responsivities in the THz range of frequencies, particularly at the plasmonic resonances at room temperature. This might provide 
the superiority of the GB-HET-based THz detectors over other plasmonic THz detectors based
on the standard heterostructures .
 Thus, the THz detectors based on the GB-HETs  can be interesting for different applications.

\section*{Acknowledgments}

The authors are grateful to D. Coquillat and F. Teppe for the information related to their
experimental data on InP HBTs operating as THz detectors. The
work was supported by the Japan Society for Promotion
of Science (Grant-in-Aid for Specially Promoted
Research  23000008) and by the Russian Scientific
Foundation (Project 14-29-00277). The works at UB
and RPI were supported by the US Air Force award 
FA9550-10-1-391 and by the US Army Research Laboratory
Cooperative Research Agreement, respectively.

\newpage


\begin{thebibliography}{9}   

\bibitem{1}
W. Mehr, J. Ch.Scheytt, J. Dabrowski, G. Lippert, Y.-H.Xie, M. C. Lemme, M. Ostling, and
G. Lupina, 
IEEE Electron Device Lett. {\bf 33}, 691, (2012)

\bibitem{2}
B. D. Kong, C. Zeng, D. K. Gaskill, K.L. Wang, K. W. Kim, Appl. Phys. Lett.
{\bf 101},  263112 (2012).

\bibitem{3}
S. Vaziri, G.Lupina, C. Henkel, A. D. Smith, M.Ostling, J. Dabrowski, G. Lippert, W. Mehr,
and M. C.Lemme 
Nano Lett. {\bf 13}, 1435 (2013).



\bibitem{4}
C. Zeng, E. B. Song, M. Wang, S. Lee, C. M. Torres,Jr., J. Tang, B. H. Weiler, and K. L.
Wang. 
Nano Lett. {\bf 13}, 2370 (2013).   

\bibitem{5}
L. Britnel, R. V. Gorbachev, R. Jalil, B.D . Belle, F.Shedin, A. Mishenko, T. Georgiou, M. I. Katsnelson, L. Eaves,
S. V. Morozov, N. M. R. Peres, J. Leist, A. K. Geim, K. S. Novoselov, and L. A. Ponomarenko,
Science, {\bf 335}, 947 (2012).



\bibitem{6}
T. Georgiou, R. Jalil, B. D. Bellee, L. Britnell, R. V. Gorbachev, S. V. Morozov, 
Y.-J. Kim, A. Cholinia, S. J. Haigh, O. Makarovsky, L. Eaves, L. A. Ponimarenko, A. K. Geim, K. S. Nonoselov, and A. Mishchenko,
Nature Nanotechnology {\bf  8}, 100 (2013). 

\bibitem{7} 
L. Britnel, R. V. Gorbachev, A. K.  Geim, L. A. Ponomarenko, A. Mishchenko, M. T. Greenaway, T. M. Fromhold, 
K. S. Novoselov,  and L. Eaves,  Nature Comm. {\bf 4}, 1794 (2013).

\bibitem{8}
M. Liu, X. Yin, and X. Zhang, Nano Lett. 12, 1482 (2012).

\bibitem{9}
C. A. Mead, Proc. IRE, {\bf 48}, 359 (1960).


\bibitem{10}
J.M. Shannon, IEE J. Solid-State Electron Devices, {\bf 3}, 142 (1979).

\bibitem{11}
M. Heiblum, D. C. Thomas, C. M. Knoedler, M. I. Nathan, Surf. Sci. {\bf 174}, 478 (1986).



\bibitem{12}
S. Luryi,
IEEE Electron Device Lett. {\bf 6}, 178 (1985).

\bibitem{13}
S. Luryi, in{\it High Speed Semiconductor Devices}, edited by S. M. Sze (Wiley, New York, 1990), p.399.

\bibitem{14} J. Xu and M. S. Shur, Double Base Hot Electron Transistor, 
United States Patent $\#$4,901,122, 
February 13 (1990).

\bibitem{15} M. S. Shur, R. Gaska, A. Bykhovski, M. A. Khan, and J. W.  Yang, 
Appl. Phys. Lett. {\bf 76}, 3298 (2000). 

\bibitem{16}
M. Asada, et.al. Jpn. J. Appl. Phys. {\bf 47}, 4375 (2008).
%


\bibitem{17}
V. Ryzhii, T. Otsuji, M. Ryzhii, V. Ya. Aleshkin, A. A. Dubinov, V. Mitin, and M.S.Shur,
J. Appl. Phys. {\bf 117}, 154504 (2015).




\bibitem{18}
V. Ryzhii,
Appl. Phys. Lett. {\bf 70}, 2532 (1997).

\bibitem{19}
V. Ryzhii,
Jpn. J. Appl. Phys. {\bf 37}, 5937 (1998).


\bibitem{20}
V. Ryzhii,  A. Satou, and T. Otsuji,
J.  Appl. Phys. {\bf 101}, 024509 (2007).

\bibitem{21}
M. I. Dyakonov and M. S. Shur,
IEEE Trans. Electron Devices, {\bf 43}, 1640 (1996).

\bibitem{22} 
W. Knap, Y. Deng, S. Rumyantsev, J.-Q. Lu, M. S. Shur, C. A. Saylor, and L. C. Brunel,
Appl. Phys. Lett. {\bf 80}, 3433 (2002).

\bibitem{23} 
X. G. Peralta, S. J. Allen, M. C. Wanke, N. E. Harff, J. A. Simmons, M. P. Lilly, J. L. Reno,
P. J. Burke, and J. P. Eisenstein, Appl. Phys. Lett. {\bf 81}, 1627 (2002).

\bibitem
{24} T. Otsuji, M. Hanabe and O. Ogawara, Appl. Phys. Lett. {\bf 85}, 2119 (2004).

\bibitem{25} 
J. Lusakowski, W. Knap, N. Dyakonova, L. Varani, J. Mateos, T. Gonzales, Y. Roelens,
S. Bullaert, A. Cappy and K. Karpierz, J. Appl. Phys. {\bf 97}, 064307 (2005).

\bibitem{26} 
F. Teppe, W. Knap, D. Veksler, M. S. Shur, A. P. Dmitriev, V. Yu. Kacharovskii, and
S. Rumyantsev, Appl. Phys. Lett. {\bf 87}, 052105 (2005).

\bibitem{27} 
V. Ryzhii, A.Satou, W.Knap, and M.S.Shur. J. Appl. Phys. {\bf 99}, 084507  (2008).

\bibitem{28} 
A. El Fatimy, F. Teppe, N. Dyakonova, W. Knap, D. Seliuta, G. Valusis, A. Shcherepetov, Y.
Roelens, S. Bollaert, A. Cappy, and S. Rumyantsev, Appl. Phys. Lett. {\bf 89}, 131926 (2006).

\bibitem{29} 
J. Torres, P. Nouvel, A. Akwaoue-Ondo, L. Chusseau, F. Teppe, A. Shcherepetov, and S. Bollaert,
Appl. Phys. Lett. {\bf 89}, 201101 (2006).

\bibitem{30}
S.A. Boubanga Tombet, Y. Tanimoto , A. Satou , T. Suemitsu , Y. Wang , H. Minamide , H. Ito, 
D. V. Fateev , V.V. Popov , and T. Otsuji,
Appl. Phys. Lett., {\bf 104},  262104 (2014).

\bibitem{31} 
Y. Kurita, G. Ducournau, D. Coquillat, A. Satou, K. Kobayashi, S.A. Boubanga-Tombet, Y.M. Meziani, V.V. Popov, W. Knap, T. Suemitsu, and T. Otsuji
Appl. Phys. Lett., {\bf 104},  251114 (2014).
  


\bibitem{32}
L. Vicarelli, M. S. Vitiello, D. Coquillat, A. Lombardo, A. C. Ferrari, W. Knap, M. Polini, 
V. Pellegrini, and A. Tredicucci,
Nature Mat. {\bf 11}, 865 (2012). 




\bibitem{33}
V. Ryzhii, T. Otsuji, M. Ryzhii, and M. S. Shur,
J. Phys. D: Appl. Phys. {\bf 45},  302001 (2012).

\bibitem{34}
V. Ryzhii, A Satou, T Otsuji, M Ryzhii, V Mitin, and M S Shur,
J. Phys. D: Appl. Phys. {\bf 46}, 315107  (2013).


\bibitem{35}
M. S. Shur {\it Physics of Semiconductor Devices,} (Prentice Hall, Mew Jersey, 1990).


\bibitem{36} 
S.~Luryi, Appl. Phys. Lett.{\bf 52}, 501 (1988).

\bibitem{37}
T. Fang, A. Konar, H. Xing, and D. Jena, Appl. Phys. Lett. {\bf 91},
092109 (2007).




\bibitem{38} 
L.A. Falkovsky and A.A. Varlamov, 
 Eur. Phys. J. B {\bf 56}, 281 (2007).

\newpage

\bibitem{39}
V.P. Gusynin, S.G. Sharapov, 
Phys. Rev. B {\bf 73}, 245411~(2006).

\bibitem{40} 
J. Cserti, Phys. Rev. B {\bf 75}, 033405 (2007).


\bibitem{41}
L. A. Falkovsky, 
J.  Phys.: Conf. Series {\bf 129},  012004 (2008).


\bibitem{42}
A. H. Castro Neto, F Guinea, N.M. R Peres, K. S. Novoselov, A. K. Geim,
Rev.  Mod. Phys. {\bf 81}  109, 2009.

\bibitem{43}
D. Svintsov, V. Vyurkov, S. Yurchenko, T. Otsuji, and V. Ryzhii,
J. Appl. Phys. {\bf 111}, 083715 (2013).

\bibitem{44}
B. N. Narozhny, I. V. Gornyi, M. Titov, M. Schutt, and A. D. Mirlin
Phys. Rev. B {\bf 91}, 035414 (2015).


\bibitem{45}
M. A. Kats, N. Y. Geneved, Z. Gaburro, and F. Capasso,
Opt. Exp. {\bf 19}, 21748 (2011). 

\bibitem{46}
L. D. Landau and E. M. Lifshitz,
{\it The Theory of Fields}, (Pergamon, Oxford, 1971).

\bibitem{47}
D. J. Griffiths, {\it Introduction to Electrodynamics}, (Benjamin Cummings, 1999)

\bibitem{48}
R. E. Colin, {it Antenna and Radiowave Propagation} (McGraw-HillNew York, 1985).

\bibitem{49}
M. Müller, J. Schmalian and L. Fritz,
 Phys. Rev. Lett.{\bf 103}, 025301 (2009).
 
 
\bibitem{50}
M. Mendoza,	H. J. Herrmann, and  S. Succi,	
Scientific Reports {\bf 3}, 1052 (2013).




\bibitem{51}
D. Coquillat, V. Nodjiadjim, A. Konczykowska, M. Riet, N. Dyakonova, C. Consejo, F. Teppe,
J. Godin, W. Knap, 
 IRMMW-THz: Int. Conf. on Infrared, Millimeter, and Terahertz Waves, Tucson, AZ, USA, 2014.

\bibitem{52}
D. Coquillat, V Nodjiadjim, A Konczykowska, N Dyakonova, C Consejo, S Ruffenach, F Teppe, M Riet, A Muraviev, A Gutin, M Shur, J Godin, and W Knap,

\bibitem{53}
   D. Coquillat, V. Nodjiadjim, A. Konczykowska, N. Dyakonova, C. Consejo, S. Ruffenach, F. Teppe, M. Riet, A. Muraviev, A. Gutin, M. Shur, J. Godin, W. Knap, 
  Proc. 40th IRMMW-THz: Int. Conf. on Infrared, Millimeter, and Terahertz Waves (Hong Kong, China), 2015.

\end{thebibliography}
\end{document}